\documentclass[aps,prl,reprint,superscriptaddress,amsmath,amssymb]{revtex4-2}
\usepackage{hyperref}
\usepackage{graphicx}
\usepackage{bm}
\usepackage{dcolumn}
\hypersetup{
    colorlinks=true,
    linkcolor=blue,
    filecolor=magenta,  
    citecolor=blue,    
    urlcolor=blue,
}
\begin{document}
\title{Reflectors Tune Near-Field Thermal Transport}

\author{Yun-Chao Hao}
\affiliation{School of Energy Science and Engineering, Harbin Institute of Technology, Harbin 150001, China}
\affiliation{Key Laboratory of Aerospace Thermophysics, Ministry of Industry and Information Technology, Harbin 150001, China}
\author{Matthias Krüger}
\email{matthias.kruger@uni-goettingen.de}
\affiliation{Institute for Theoretical Physics, University of Göttingen, 37077 Göttingen, Germany}
\author{Mauro Antezza}
\affiliation{Laboratoire Charles Coulomb(L2C), UMR 5221 CNRS-Universitéde Montpellier, F-34095 Montpellier, France}
\affiliation{Institut Universitaire de France, 1 rue Descartes, F-75231 Paris Cedex 05, France}
\author{Cheng-Long Zhou}
\affiliation{School of Energy Science and Engineering, Harbin Institute of Technology, Harbin 150001, China}
\affiliation{Key Laboratory of Aerospace Thermophysics, Ministry of Industry and Information Technology, Harbin 150001, China}
\author{Hong-Liang Yi}
\affiliation{School of Energy Science and Engineering, Harbin Institute of Technology, Harbin 150001, China}
\affiliation{Key Laboratory of Aerospace Thermophysics, Ministry of Industry and Information Technology, Harbin 150001, China}
\author{Yong Zhang}
\email{yong\_zhang@hit.edu.cn}
\affiliation{School of Energy Science and Engineering, Harbin Institute of Technology, Harbin 150001, China}
\affiliation{Key Laboratory of Aerospace Thermophysics, Ministry of Industry and Information Technology, Harbin 150001, China}
\date{\today}

\begin{abstract}
We explore near-field thermal radiation transport in nanoparticles embedded within a multilayer slab structure, focusing on dynamic modulation of heat flux via cavity interactions. Our findings reveal that by tuning the distance between reflectors and nanoparticles, thermal transport can be significantly suppressed or enhanced, driven by selective excitation of surface modes within the cavity. By precisely adjusting inter-slab gaps, we achieve multi-order control over thermal flux while maintaining stability across a broad range of configurations. Notably, internal slab arrangement plays a pivotal role, with compact designs yielding the most pronounced effects. This work unveils a novel mechanism for manipulating near-field heat transfer, with exciting potential for nanoscale thermal management and thermal sensing technologies.
\end{abstract}
\maketitle

Advancements in nanotechnology and near-field radiative heat transfer (NFRHT) have 
made high-sensitivity mK-level temperature detection essential for microelectromechanical 
systems \cite{lang_2017_dynamic, luo_2024_observation, svendagebiehs_2021_nearfield,Bimonte2017}, including  slab structures \cite{fiorino_2018_giant, fiorino_2018_a, rincngarca_2022_enhancement, rohithmittapally_2023_quantifying, rohithmittapally_2023_probing} and spherical structures \cite{shen_2023_surface, song_2015_enhancement, shi_2015_nearfield, lucchesi_2021_nearfield}. As efficient thermal feedback platform \cite{vassiliosyannopapas_2013_spatiotemporal, zograf_2017_resonant, artyomassadillayev_2021_thermal, zhu_2016_temperaturefeedback}, 
nanoparticles exhibit local thermal enhancements in multi-particle arrangements \cite{benabdallah_2011_manybody, philippebenabdallah_2019_multitip, thanhxuanhoang_2024_photonic, Muller2017}, suggesting their suitability for near-field thermal information relay,  bolstered by levitated nanoparticles \cite{agrenius_2023_interaction}. However, NFRHT efficiency in particle structures is dampened by the distance-proportional law $d^{-6}$ \cite{narayanaswamy_2008_thermal} (where $d$ represents the separation distance), and in slab structures by $d^{-2}$ \cite{volokitin_2004_resonant}, indicating challenges in long-range thermal transport. Introducing  slabs between particles can provide additional thermal exchange channels 
\cite{zhang_2023_enhanced, zhang_2019_giant, fang_2023_nearfield}, akin to the enhancement 
effects seen in multi-layer slab structures \cite{lim_2020_surfaceplasmonenhanced, messina_2012_threebody, philippebenabdallah_2014_nearfield, iizuka_2018_significant}, potentially supported by the $d^{-3}$  law that particle and slab structures follow as a function of distance $d$  \cite{mulet_2001_nanoscale}. Incorporating intermediate slabs significantly boosts radiative heat transfer efficiency in nanoparticle assemblies but offers limited tunability \cite{zhang_2023_enhanced, zhang_2019_giant, fang_2023_nearfield}. 

\begin{figure}[t]
\includegraphics[width=0.48\textwidth]{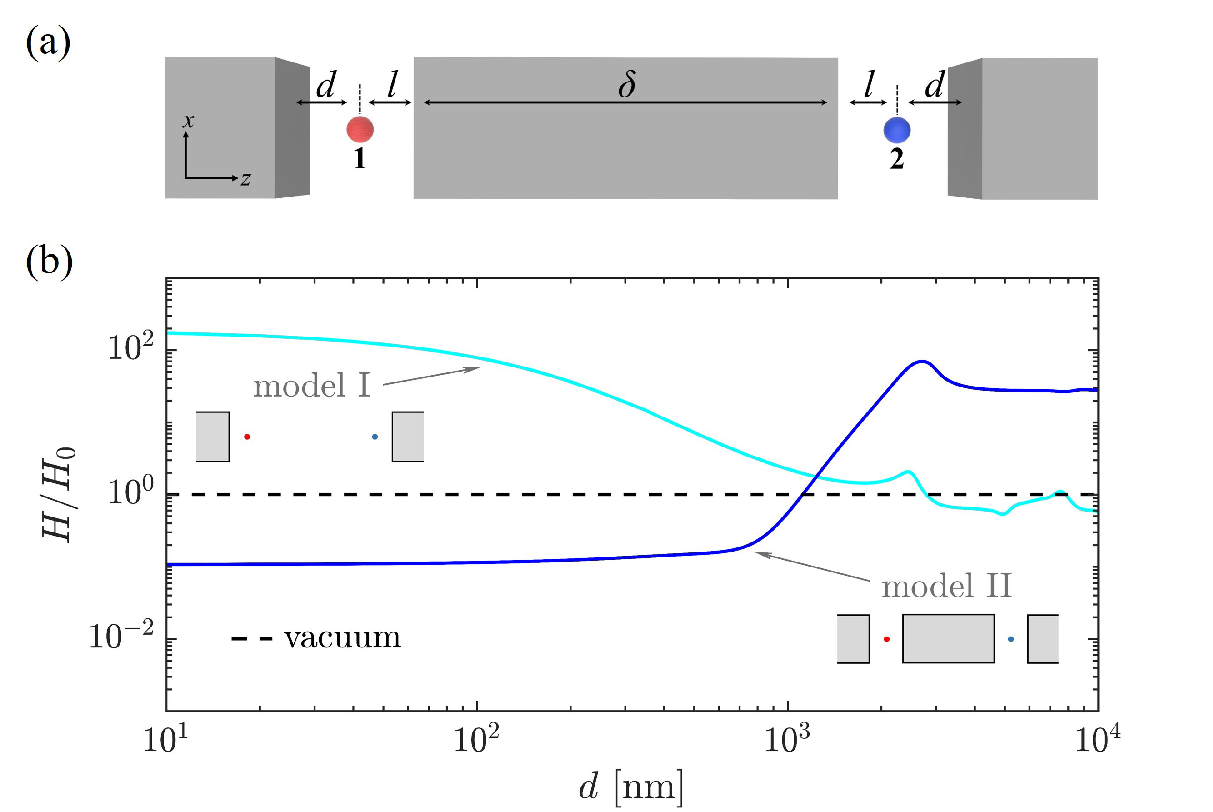}
\caption{\label{fig1}(a) Diagram illustrating near-field thermal transport between particles 1 and 2, mediated by slabs. The 'repeater' is the intermediate slab with thickness $\delta$, and the 'reflectors' are the adjacent semi-infinite slabs, with their distances from the particles labeled $l$ and $d$, respectively. (b) Compares the thermal conductence $H$ of different systems, normalized to vacuum conductence $H_{0}$, as functions of the distance $d$ between the reflectors and particles. Models I (only reflectors) and II (repeater and reflectors) are shown.}
\end{figure}

The study of electromagnetic wave interactions with matter in a cavity has greatly advanced fields like cavity quantum electrodynamics \cite{friskkockum_2019_ultrastrong, giacomojarc_2023_cavitymediated} and cavity optomechanics \cite{aspelmeyer_2014_cavity, delosrossommer_2021_strong}. In the domain of NFRHT, substantial research has been devoted to controlling heat flux by adjusting the gap with the cavity acting as a thermal source \cite{thompson_2019_nanoscale, messina_2014_threebody, khayam_2024_enhancement}. However, the use of cavity modes as a mechanism to modulate heat exchange of substances (typically nanoparticles) within the cavity has not yet been studied. Here, we introduce reflective slabs {\it on the outside} of the nanoparticles to form a cavity structure, allowing us to control the radiative properties by tuning the cavity parameters [see Fig.\thinspace\ref{fig1}(a)]. A central slab of thickness $\delta$, termed 'repeater', is shown to greatly enhance thermal flux when outer (semi-infinite) surfaces, termed 'reflectors', are at a large distance. Depending on distance, these reflectors select the dominant surface modes (SPhPs) \cite{mulet_2002_enhanced, heinisch_2012_mie}, thereby tuning the transfer between the nanoparticles. Particles and slabs are made of non-magnetic dielectric materials. This thus introduces a novel method of selecting modes and thereby tuning near-field transfer. 

The spatial arrangement is shown in Fig.\thinspace\ref{fig1}(a) with the nanoparticles having distances $l$ and $d$ from the repeater and reflector, respectively. With the entire setup at temperature $T$ and particle 1 warmed to $T+\Delta T$ (where $T\gg\Delta T$), $\Delta T$ causes radiative energy exchange, where we here focus on transfer $\Phi$ from particle 1 to particle 2, with $\Delta T>0$ \cite{svendagebiehs_2010_mesoscopic, francoeur_2010_local}. For particle radius $R \ll \lambda_{T}$ (thermal wavelength $\lambda_{T} = c\hbar/k_{B}T$) and $R$ small compared to $l$ and $d$, particles are approximated as dipoles \cite{lukasnovotny_2006_principles}. Accordingly, the interparticle thermal conductence $H = \partial \Phi /\partial T$ follows  \cite{kirylasheichyk_2022_radiative, kirylasheichyk_2017_heat,benabdallah_2011_manybody}
\begin{eqnarray}
H=64\pi^2\int_0^{\infty} \frac{d \omega}{2 \pi} \hbar \omega n^{\prime} k_0^4\operatorname{Im}(\alpha_{1})\operatorname{Im}(\alpha_{2}) \operatorname{Tr}\left(\mathbb{G} \mathbb{G}^{\dagger}\right),\label{Eq1}
\end{eqnarray}
\noindent
where $\omega$ and $k_{0}=\omega/c$ respectively represent the angular frequency and the free space wave number. $n^{\prime}$ is the derivative of the Bose-Einstein distribution $n=[\mathrm{exp}(\hbar\omega/k_{B}T)-1]^{-1}$ with respect to $T$. $\alpha_{i}(\omega)= R_{i}^3[\varepsilon_{i}(\omega)-1] /[\varepsilon_{i}(\omega)+2]$ is the particles' electrical polarizability with $\varepsilon(\omega)$ the dielectric permittivity \cite{albaladejo_2010_radiative}. 

$\mathbb{G}$ refers to the dyadic Green's function (DGF), which characterizes the electromagnetic interaction between two particles. In the Fourier-transformed two-dimensional wavevector space, due to the symmetry of the transverse structure, we obtain
\begin{eqnarray}
\mathbb{G}=\int_0^{\infty}\frac{d k_\rho}{2 \pi} k_\rho g,\label{Eq2}
\end{eqnarray}
\noindent
where $k_{\rho}$ is the transverse wave number, with the relationship $k_{z}=\sqrt{k_{0}^{2}-k_{\rho}^{2}}$ and $g$ is given by
\begin{eqnarray}
g=\frac{i}{4 k_z k_0^2}\left(\begin{array}{ccc}
k_{z}^{2}\alpha+k_{0}^{2}\beta & 0 & 0 \\
0 & k_{z}^{2}\alpha+k_{0}^{2}\beta & 0 \\
0 & 0 & 2k_{\rho}^{2}\gamma
\end{array}\right).\label{Eq3}
\end{eqnarray}
\noindent
Here we introduced abbreviations 
\begin{eqnarray}
\begin{aligned}
\alpha &= (A^{TM}-C^{TM}) e^{i k_z l}+(D^{TM}-B^{TM}) e^{-i k_z l} \\
\beta &= (A^{TE}\,+C^{TE}\,) e^{i k_z l}+(D^{TE}\,+B^{TE}\,) e^{-i k_z l} \\
\gamma &= (A^{TM}+C^{TM}) e^{i k_z l}+(D^{TM}+B^{TM}) e^{-i k_z l}
\end{aligned},\label{Eq4}
\end{eqnarray}

\noindent
where coefficients $A-D$ denote the field amplitudes constituting the Weyl components of DGF. The superscripts TM and TE denote the polarization (see supplementary materials \cite{supplementary_materials} for details).

To simplify the analysis, we fix $T=300\, \mathrm{K}$, $l=100\, \mathrm{nm}$, $\delta=3\, \mathrm{\mu m}$, and $R_{1}=R_{2}=5\, \mathrm{nm}$, and vary the distance $d$. The impact of $\delta$ is  discussed in the supplementary materials \cite{supplementary_materials}. 
The material properties of the particles and slabs are chosen carefully: The resonance frequency for surface localized modes of the paricle are determined by the polarizability condition $\varepsilon(\omega_{np})+2=0$. The chosen isotropic SiC particles \cite{palik_2003_handbook} show that $\omega_{np}$ is $1.756\times 10^{14}\, \mathrm{rad\ s^{-1}}$. For the slabs, the SPhPs resonance, defined by $\varepsilon(\omega_{sp})+1=0$ \cite{joulain_2005_surface}, yields $\omega_{sp} = 1.787\times 10^{14}\, \mathrm{rad\ s^{-1}}$ for SiC, which does not perfectly match with $\omega_{np}$; surface mode coupling between particles and slabs of identical material invariably leads to monochromaticity loss \cite{dong_2018_longdistance, fang_2023_active}. To address this, we exploit the Mie resonance effect \cite{francoeur_2011_electric, zhao_2009_mie} to select slab materials with $\omega_{sp}$ closely matching that of SiC particles, i.e., $\omega_{sp} \simeq \omega_{np}$ (for  details see the supplementary materials \cite{supplementary_materials}). Adjustments to $\omega_{sp}$, to finely tune resonance without altering peak values, are explored through isotope effects \cite{papadakis_2021_deepsubwavelength, xie_2023_isotope}, photonic crystal structures \cite{zhang_2019_giant, fang_2023_nearfield}, among other strategies \cite{messina_2012_threebody}.

The modification of an emitter's electromagnetic environment can enhance or suppress its interaction with light, a phenomenon known as the Purcell effect. By engineering the electromagnetic environment (e.g., through cavity design), one can control the coupling strength. We here study the influence of  reflectors along the $z$-axis, with Fig.\thinspace\ref{fig1}(b) showing thermal transport as a function of distance $d$. To understand the influence of the interior slab, we consider a reflectors-only configuration (model I) and contrast it with the setup depicted in Fig.\thinspace\ref{fig1}(a) (model II). In model I, when the reflectors are distant from the particles ($d=10\,\mathrm{\mu m}$), evanescent modes become ineffective when crossing the vacuum, causing $H$ to approach $H_{0}$, the value found for two nanoparticles in empty space. The oscillations observed are due to the mutual reflection of propagating photons between slabs. As the reflectors approach the particles , i.e., with decreasing $d$, $H$ strongly increases, by two orders of magnitude.  We attribute this enhancement to surface modes (SPhPs), highlighting the reflectors' role in boosting particle thermal transport, consistent with related research findings \cite{kirylasheichyk_2018_heat, kirylasheichyk_2023_longrange, zhang_2021_manyparticle}. In model II, with the reflectors  distant from the particles, photon-induced oscillations are minimal relative to the enhancement from the repeater's surface modes, making the curve essentially independent of $d$. This effectively equates to having only the repeater involved in heat exchange, with its enhancement effect aligning with related research findings \cite{zhang_2019_giant,fang_2023_nearfield,zhang_2023_enhanced}. Interestingly, moving the reflectors closer reduces $H$ by three orders of magnitude, even below $H_{0}$, which suggests potential thermal flow suppression in compact configurations and challenges conventional understanding \cite{shen_2009_surface, benabdallah_2013_heat}. This configuration-selective thermal channeling suggests that near-field thermal management devices, like thermal switches, can robustly control thermal flux through cavity gap adjustments, without moving the heat source. Notably, within the approximate ranges of $d \alt l$ or $d\agt\delta$, each configuration exhibits thermal flux nearly independent of $d$, akin to radiative heat transfer saturation under multi-body effects \cite{latella_2020_saturation}. This implies that controlling thermal flux between particles by altering cavity gaps is largely robust.

Coupling strength shifts between particle and slab structure during thermal exchange depend on cavity mode and particle resonance frequency matching. The behavior of $H$ with respect to $d$ as shown in Fig.\thinspace\ref{fig1}(b) is elucidated by examining the dispersion in momentum space. We therefore consider  $h(\omega,k_{\rho})$ to characterize the distribution of $H$ in momentum space and frequency domain, given by
\begin{equation}
h=64\pi^2\hbar\omega n^{\prime}k_{0}^{4}\operatorname{Im}(\alpha_{1})\operatorname{Im}(\alpha_{2})k_{\rho}\operatorname{Tr}\left(Gg^{\dagger}+gG^{\dagger}\right) \label{Eq5}
\end{equation}
\noindent
with $H=\int_0^\infty \frac{d\omega}{2\pi} \int_0^\infty \frac{dk_\rho}{2\pi} h(\omega,k_\rho)$ and
\begin{equation}
G=\int_0^{k_{\rho}} \frac{d k_\rho}{2 \pi} k_\rho g.\label{Eq6}
\end{equation}

\begin{figure}[t]
\includegraphics[width=0.48\textwidth]{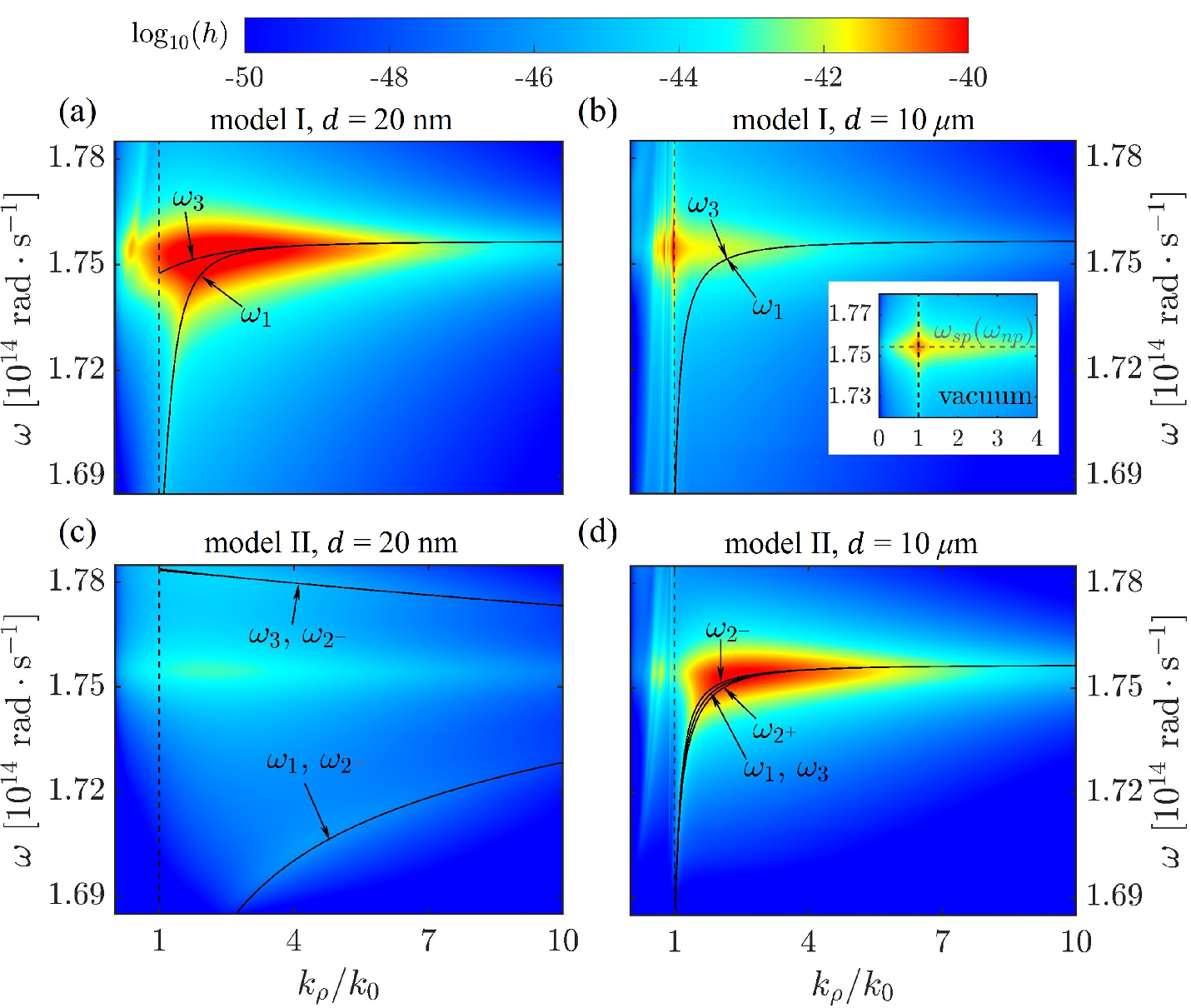}
\caption{\label{fig2}$h$ maps for models I and II at different $d$. Solid lines indicate system's cooperative surface modes dispersion curves; dotted lines separate free-space propagating ($k_{\rho} < k_{0}$) and evanescent ($k_{\rho} > k_{0}$) modes. Inset illustrates $h$ map in vacuum.}
\end{figure}

Fig.\thinspace\ref{fig2}(a) and (b) illustrate $h$ as function of $\omega$ and $k_\rho$ for model I at $d=20\,\mathrm{nm}$ and $10\,\mathrm{\mu m}$, respectively. Due to the overwhelming thermal flux provided around the above mentioned resonance frequencies \cite{svendagebiehs_2016_nearfield}, we focus on the range of frequencies around them. In this region, SPhPs are excited solely by TM polarized waves \cite{latella_2017_radiative}, $\beta \simeq 0$ in Eq.\thinspace(\ref{Eq3}), and we consider the dispersion curves in the evanescent mode region that are derived from the Fabry-Pérot-like denominator term 
\begin{equation}
1-R_{+}^{TM} R_{-}^{TM} e^{2 i k_{z}(l+d)}=0.\label{Eq7}
\end{equation}
\noindent
$R_{+}^{TM}$ and $R_{-}^{TM}$ are the reflection coefficients at the interfaces of the repeater and reflectors within the vacuum gap surrounding the particles, respectively (see the supplementary materials for details \cite{supplementary_materials}).

When the reflectors are close to the particles ($d=20\,\mathrm{nm}$), the optical distance between their interfaces approaches the inter-particle distance. The direct thermal emission and SPhPs received by particle 2 have comparable magnitudes. Under the coherent action of the reflectors' surface modes, the dispersion lines manifest as two symmetric and antisymmetric dispersion lines, denoted as $\omega_{1}$ and $\omega_{3}$. These lines tend towards degeneracy with increasing $k_{\rho}$ and reducing penetration depth \cite{francoeur_2010_local, svendagebiehs_2016_nearfield}. By comparing with the $h$ distribution in vacuum shown in the inset, Fig.\thinspace\ref{fig2}(b), one can observe a distinct bright band near the dispersion curves, indicative of monochromatic enhancement at the $\omega_{sp}$. When the reflectors are distant from the particles ($d=10\,\mathrm{\mu m}$), the surface modes can no longer couple due to their inability to tunnel through the vacuum gap, resulting in $\omega_{1}$ and $\omega_{3}$ degenerating into the dispersion curve of a single interface. Here, the magnitude of SPhPs received by particle 2 is markedly less than direct thermal emissions from particle 1, with multiple bright bands formed by Fabry-Pérot-like standing waves emerging in the propagating mode region \cite{latella_2017_radiative}, correlating with fluctuations observed in Fig.\thinspace\ref{fig1}(b).
\begin{figure}[b]
\includegraphics[width=0.48\textwidth]{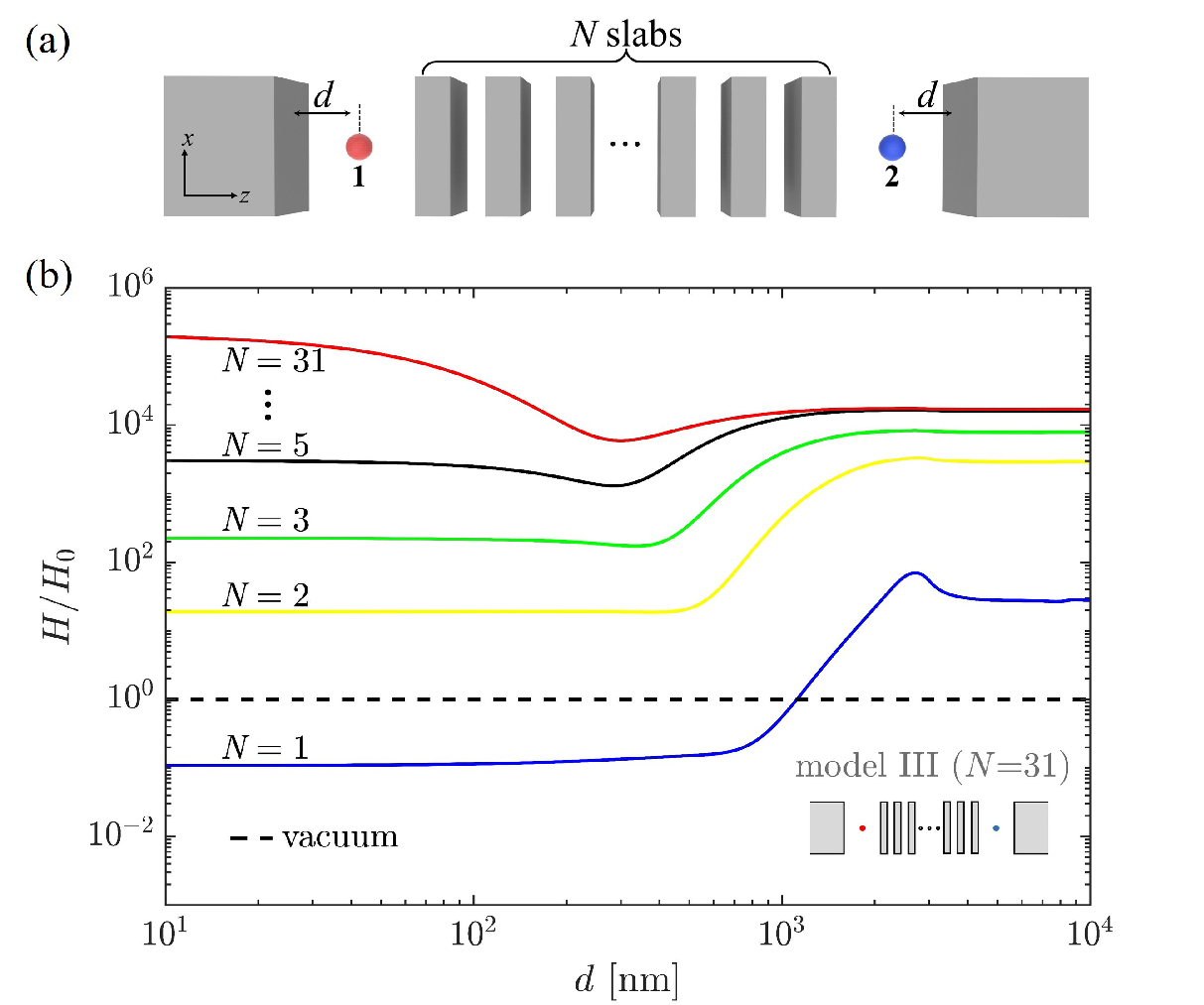}
\caption{\label{fig3} System with a periodic multilayer repeater consisting of $N$ slabs and $N-1$ vacuum layers, each with a thickness of $\delta/(2N-1)$. (b) Shows how heat transfer coefficient $H$ varies with distance $d$ between the reflectors and particles across different $N$ configurations.}
\end{figure}

Fig.\thinspace\ref{fig2}(c) and (d) discuss model II under the same conditions, where the addition of a single-layer repeater introduces two new dispersion lines, denoted $\omega_{2^{+}}$ and $\omega_{2^{-}}$. Close proximity between reflectors and particles extends symmetric and antisymmetric dispersion lines over a broader frequency due to surface modes coupling through the small vacuum gap, similar to Rabi splitting in quantum cavity \cite{chikkaraddy_2016_singlemolecule, santhosh_2016_vacuum}. Moreover, due to the repeater's sufficient thickness, $\omega_{1}$ and $\omega_{2^{+}}$, $\omega_{3}$ and $\omega_{2^{-}}$ become degenerate. Behind this structural effect lies a wavenumber selective mechanism, monochromatic thermal emission from particle 1 spreads across the entire strong reflection band, while high-$k_{\rho}$ thermal channels close, hindering surface mode coupling within the repeater and suppressing monochromatic emission (for specific details, see the supplementary materials \cite{supplementary_materials}). This reveals the high tunability of thermal transport in model II and implies that adding more slabs does not inherently boost thermal radiation transport \cite{ilic_2018_active}. When the reflectors are distant, the coherent action within the vacuum gap vanishes, causing the surface mode dispersion lines $\omega_{1}$ and $\omega_{3}$ supported by the reflectors to degenerate back into a single interface dispersion, and $\omega_{2^{+}}$ and $\omega_{2^{-}}$ supported by the repeater to recouple into a single thin-film dispersion, showing thermal flux enhancement akin to that seen in Fig.\thinspace\ref{fig2}(a).

\begin{figure}[b]
\includegraphics[width=0.41\textwidth]{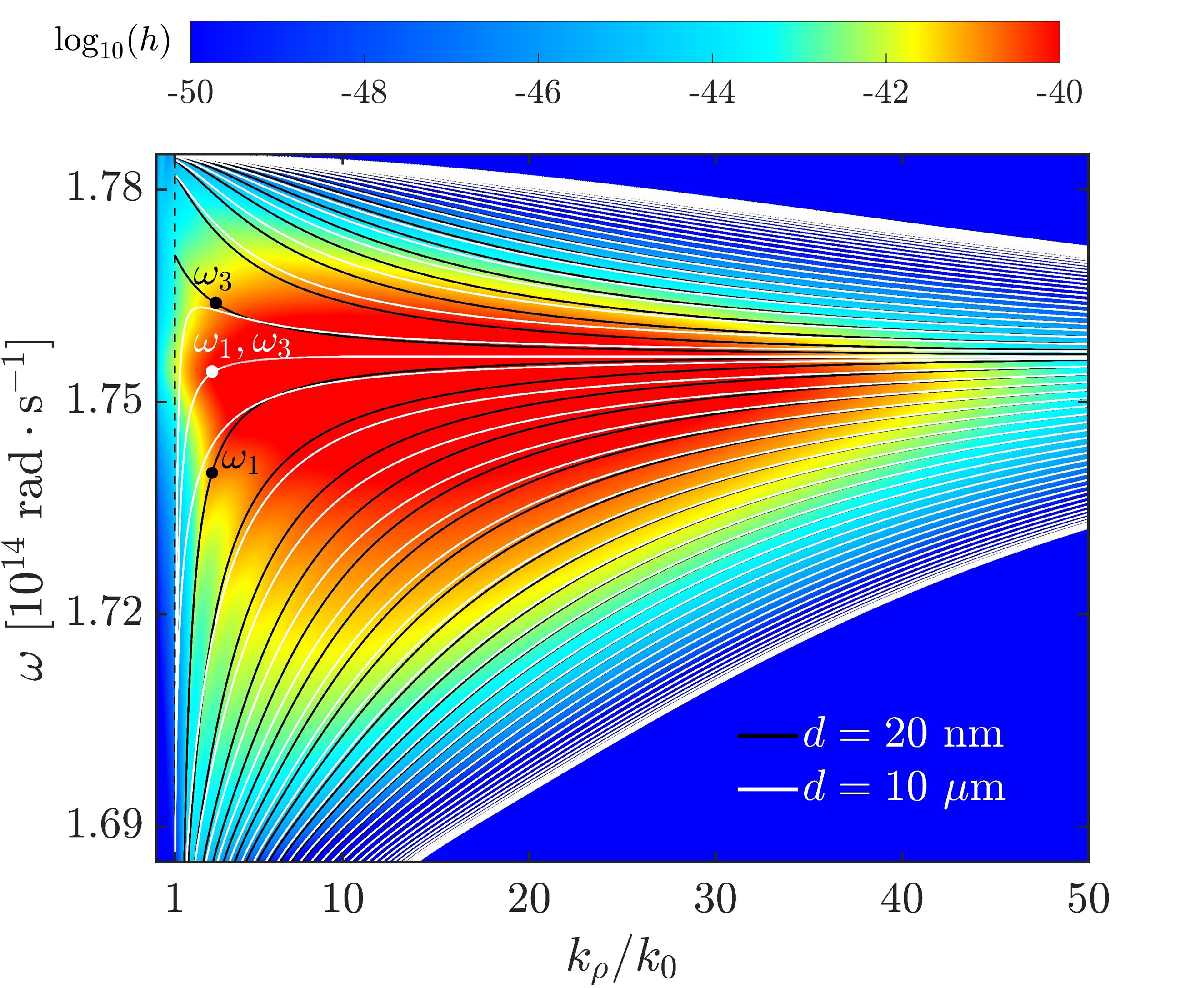}
\caption{\label{fig4}The $h$ map for model III at $d=20\,\mathrm{nm}$ shows coupled surface wave dispersion lines. Black and white solid lines represent the cases for $d=20\,\mathrm{nm}$ and $d=10\,\mathrm{\mu m}$, respectively. Dispersion lines $\omega_{1}$ and $\omega_{3}$ are introduced by reflectors, with the remaining solid lines introduced by the multilayer repeater.}
\end{figure}

Clearly, for a single-layer repeater, sufficient thickness ensures the distance of energy transmission but limits the tunneling of high-$k_{\rho}$ evanescent modes. Considering this, we explored a periodic multilayer repeater interspersed with vacuum layers [see Fig.\thinspace\ref{fig3}(a)] to enhance surface electromagnetic mode utilization. This multilayer repeater, with slabs count $N$ and total thickness $\delta$, essentially acts as a hyperbolic metamaterial (HMM), supporting coherent thermal transport of broad bandwidth hyperbolic modes over several magnitudes greater penetration depth than surface modes \cite{lang_2014_large, biehs_2012_hyperbolic}. Fig.\thinspace\ref{fig3}(b) illustrates $H$'s dependence on $d$ and $N$. Considering the impact on material properties, we limit $N$ to a maximum of 31, designating the system as model III. When the reflectors are distant from the particles, $H$ surpasses model II, benefiting from the multilayer's hyperbolic modes \cite{zhang_2019_giant, fang_2023_nearfield}. As the reflectors approach the particles, $H$ increases by an order of magnitude, underscoring reflectors' role in boosting thermal transfer. Increasing $N$ effects a transition from the blue to the red line in Fig.\thinspace\ref{fig3}(b), indicating that the configuration of the repeater alone can significantly increase thermal flow, and the joint action of the repeater and reflectors can further extend the modulation range, even suppressing thermal flow. This demonstrates the system's capability to customize thermal flow based on the internal structure of the repeater and the position of the reflectors, enabling enhanced or suppressed thermal flow modulation in compact configurations without compromising system robustness. It opens promising pathways for thermal management and other fields.

Model III demonstrates enhanced thermal flow with closer reflectors compared to model II, as shown by the $h$ distribution for model III at $d=20\,\mathrm{nm}$ in Fig.\thinspace\ref{fig4}, showing a significant improvement in penetration depth over models I and II. This enhancement stems from the coupling of short-range SPhPs at the HMM interface, facilitating high-$k_{\rho}$ thermal transport. Bragg scattering within the periodic photonic crystal alternately strengthens or weakens wave coherence, creating photonic band gaps. The $2(N+1)$ solid lines based on Eq.\thinspace(\ref{Eq7}) represent the dispersion curves of cooperative surface modes, where $\omega_{1}$ and $\omega_{3}$ are the two dispersion lines introduced by the reflectors, and the remaining $2N$ dispersion lines are introduced by the repeater. The black and white lines depict the cases for $d=20\,\mathrm{nm}$ and $d=10\,\mathrm{\mu m}$, respectively. Is is evident that the black lines are pushed to higher-$k_{\rho}$ regions due to the coupling of surface modes within the vacuum gap. Hence, as the reflectors approach the particles, the outward expansion of the dispersion lines supported by the multilayer repeater leads to monochromatic loss in thermal flux, which is compensated by the introduction of dispersion lines supported by the reflectors. This explains the fluctuating behaviour of the red curve in Fig.\thinspace\ref{fig3}(b). Despite model III featuring multiple thermal exchange channels, each channel's thermal transfer probability is much smaller than model II's single channel (for specific details, see the supplementary materials \cite{supplementary_materials}). The enhancement of thermal flux is fundamentally due to the increase in penetration depth.

In summary, we discover novel tunability  of transfer through  multilayer slab structures via controlling the dominant transverse wavevectors using outside reflector surfaces. We showed how this can be used to tailor transfer by matching or de-matching modes in various geometries, including single or multilayer configurations with robust properties. The mechanical motion of the cavity can be achieved using a nanoscale displacement platform or an optomechanical cavity \cite{aspelmeyer_2014_cavity}, and even extended to develop nanoscale mechanical vibration sensing systems driven by thermal information. The impact of the single-side reflector's motion is detailed in the supplementary materials \cite{supplementary_materials}. This has promising implications for micro/nanoscale thermal radiation \cite{chapuis_2023_thermal, pascale_2023_perspective}, levitation dynamics \cite{rieser_2022_tunable, vijayan_2024_cavitymediated}, local responses in sub-micron gaps \cite{hakansalihoglu_2023_nonlocal, svendagebiehs_2023_far}, thermal relaxation in compact nanostructures \cite{philippebenabdallah_2022_controlling, sabiehs_2022_heat}, and Casimir interactions between nanoparticles and slabs \cite{xu_2022_observation, matthiaskrger_2011_nonequilibrium}, among others.

\begin{acknowledgments}
This work was supported by the National Natural Science Foundation of China (Grant No. 52076056, No. U22A20210). M.K. acknowledges the support from the Volkswagen Foundation through SPRUNG. M.A. acknowledges the grant ``CAT'', No. A-HKUST604/20, from the ANR/RGC Joint Research Scheme sponsored by the French National Research Agency (ANR) and the Research Grants Council (RGC) of the Hong Kong Special Administrative Region.
\end{acknowledgments}

\end{document}